\documentclass[letterpaper,twocolumn,aps,showpacs,12point,prl]{revtex4-1}

\usepackage{amsmath}
\usepackage{amssymb}

\usepackage{graphicx}% Include figure files
\usepackage{mathtools}
\usepackage{color}
\usepackage{epstopdf}
\usepackage{bm}% bold math

\newcommand{\ket}[1]{|#1\rangle}

\usepackage{graphicx}
\begin{document}

%\preprint{APS/123-QED}

\title{Coherence and Rydberg blockade of atomic ensemble qubits}

\author{{M. Ebert}}%
 \email{mebert@wisc.edu}
\author{{M. Kwon}}
\author{{T. G. Walker}}%
\author{{M. Saffman}}%

\affiliation{%
Department of Physics, University of Wisconsin, 1150 University Avenue, Madison, Wisconsin 53706, USA    
}%

\date{\today}

\begin{abstract}
	We demonstrate $\ket{W}$ state encoding of multi-atom ensemble qubits.  Using optically trapped Rb atoms the $T_2$ coherence time  is 2.6(3) ms for  $\bar N=7.6$ atoms and scales approximately inversely with the number of atoms. Strong Rydberg blockade between two ensemble qubits is demonstrated with a fidelity of $0.89(1)$ and a fidelity of  $\sim \hspace{-.05cm}1.0$ when postselected on control ensemble excitation. These results are a significant step towards deterministic entanglement of atomic ensembles. 
\end{abstract}


\maketitle

%intro

Qubits encoded in hyperfine states of neutral atoms
are a promising approach for scalable implementation
of quantum information processing\cite{Ladd2010}. While a qubit can be encoded in a pair of ground states of a single atom, it is also possible to encode a qubit, or even multiple qubits, in an $N$ atom ensemble by using Rydberg blockade to enforce single excitation of one of the qubit states\cite{Lukin2001,Brion2007d}. Ensemble qubits have several interesting features in comparison to single atom qubits. Using an array of traps it is simpler to prepare many ensemble qubits with $N\ge 1$ for each ensemble, than it is to prepare an array with exactly one atom in each trap which remains an outstanding challenge\cite{Bakr2009, *Sherson2010,Carpentier2013,Piotrowicz2013, *Nogrette2014}.  In addition,  a $\ket{W}$ state ensemble qubit encoding is maximally robust against loss of a single atom\cite{Dur2000}, which can be remedied with error correction protocols\cite{Brion2008},  while atom loss is a critical error for single atom qubits. Furthermore an ensemble encoding facilitates strong coupling between atoms and light, an essential ingredient for quantum networking protocols\cite{Duan2001,*Sangouard2011} and atomic control of photonic interactions in Rydberg blockaded ensembles\cite{Peyronel2012,*Parigi2012,*Maxwell2013,*Gorniaczyk2014,*Tiarks2014}.  As the atom-light coupling strength grows with the number of atoms,  
recent experiments\cite{Peyronel2012,*Parigi2012,*Maxwell2013,*Gorniaczyk2014,*Tiarks2014},\cite{Baur2014} and theory proposals\cite{Paredes-Barato2014,*He2014,*Khazali2015} are based on  ensembles with $N>100$. We are focused here on studying the physics of ensembles for computational qubits and therefore work with smaller ensembles with up to $N\sim 10$ atoms. 

In this letter we demonstrate and study the coherence and interactions of atomic ensemble qubits. We measure the $T_2$ coherence time of ensemble qubits achieving a ratio of coherence time to single qubit $\pi$ rotation time of $\sim 2600$. We furthermore  proceed to demonstrate strong Rydberg blockade between two, spatially separated ensemble qubits.  Together with the recent demonstration of entanglement between a Rydberg excited ensemble and a propagating photon\cite{LLi2013} these results establish a path towards both local and remote entanglement of arrays of ensemble qubits, which will enable enhanced quantum repeater architectures\cite{Han2010,*Zhao2010,*Brion2012}.

The computational basis states of the ensemble qubits are
\begin{equation}
    \ket{\bar 0} = |0_{1} ... 0_{N}\rangle,~~  \ket{\bar 1} = \frac{1}{\sqrt N}\sum_{j=1}^N |0_{1}0_{2}... 1_j ... 0_{N}\rangle,
 \label{eqn_qubitBasis}
\end{equation}
where $\ket{0_j}$ and $\ket{1_j}$ are two ground states of the $j^{\rm th}$ atom in an $N$ atom sample\cite{2015_ensemble_note1}. 
The state $\ket{\bar 1}$, which is a symmetric superposition of one of the $N$ atoms being excited,  is commonly referred to as a $\ket{W}$ state in the quantum information literature. 

Gate protocols for 
ensemble qubits  differ slightly from the single atom qubit case \cite{Jaksch2000,Lukin2001} as all operations must use blockade to prohibit multi-atom excitation.
Gate operations are performed via the collective, singly excited Rydberg state 
$$
\ket{\bar r} = \frac{1}{\sqrt N}\sum_{j=1}^N |0_{1}0_{2}... r_j ... 0_{N}\rangle,
$$
where $\ket{r_j}$ is the Rydberg state of the $j^{\rm th}$ atom. 
A single qubit rotation $R(\theta,\phi)$ with area $\theta$ and phase $\phi$  between ensemble states $\ket{\bar 0}, \ket{\bar 1}$ is implemented as the three pulse sequence 
$\ket{\bar 1} \xrightarrow[\pi]{\Omega} \ket{\bar r},$ 
$\ket{\bar r} \xleftrightarrow[R(\theta,\phi)]{\Omega_N} \ket{\bar 0},$
$\ket{\bar r} \xrightarrow[\pi]{\Omega} \ket{\bar 1}$. 
Note that the coupling strength between states $\ket{\bar 1}, \ket{\bar r}$ is  the single atom Rabi frequency $\Omega$ while the coupling between 
$\ket{\bar 0}, \ket{\bar r}$ is at the collective Rabi frequency  $\Omega_N=\sqrt N \Omega.$ Since $\Omega_N$ depends on $N$, the one-qubit gate pulse lengths  depend on the number of atoms. A $C_Z$ gate between control and target ensembles $\rm c,t$ is implemented as the three pulse sequence 
$\ket{\bar 1}_{\rm c} \xrightarrow[\pi]{\Omega} \ket{\bar r}_{\rm c},$ 
$\ket{\bar 1}_{\rm t} \xleftrightarrow[2\pi]{\Omega} \ket{\bar r}_{\rm t},$ 
$\ket{\bar r}_{\rm c} \xrightarrow[\pi]{\Omega} \ket{\bar 1}_{\rm c}$. The $C_Z$ gate pulses do not depend on the number of atoms. 
The $N$ dependence of the one-qubit gates can be strongly suppressed using adiabatic pulse sequences so that high fidelity gate operations are possible with small, but unknown values of $N$\cite{Beterov2013a}.

\begin{figure}[!t]
  \includegraphics[width=0.99\columnwidth]{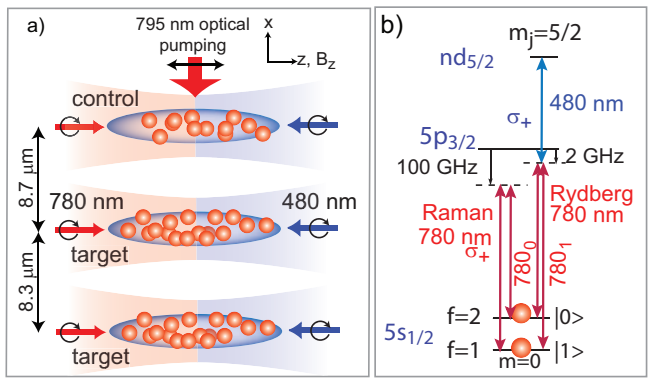}
\vspace{-.6cm}
  \caption{(color online)
    Experimental geometry a) and transitions used for qubit control b). The Raman light is only used for preparation of product states, as discussed in connection with Fig. \ref{fig_WStateT2}.  }
  \label{fig_maindia}
\end{figure}

The experimental setting is as described in \cite{Ebert2014}. In brief we  prepare a cold sample of $^{87}$Rb atoms in a magneto-optical trap (MOT) and then load a variable number of atoms into optical dipole traps. The dipole traps shown in Fig. \ref{fig_maindia} are formed by focusing 1064 nm light to waists ($1/e^2$ intensity radii) of  $3.0 ~\mu\rm m$. The atoms are cooled to a temperature of $\sim 150~\mu\rm K$ in 1-1.5 mK deep optical potentials. This gives approximately Gaussian shaped density distributions with typical standard deviations  $\sigma_\perp= 0.7~\mu \rm m$ perpendicular to the long trap axis and $\sigma_z=7~\mu\rm m$ parallel to the long axis.  The estimated density at trap center is $n/N= 5\times 10^{16} ~\rm m^{-3}$. We apply a bias magnetic field along the trap axis of $B_z=0.24~\rm  mT$ and  optically pump into $\ket{0}\equiv\ket{5s_{1/2},f=2,m_f=0}$ using $\pi$ polarized 795 nm light resonant with $\ket{5s_{1/2},f=2} \rightarrow \ket{5p_{1/2},f=2}$ and 780 nm repump light resonant with 
$\ket{5s_{1/2},f=1} \rightarrow \ket{5p_{3/2},f=2}$. 

Rydberg excitation coupling $\ket{\bar 0}, \ket{\bar r}$ is performed by off-resonant two-photon transitions via $5p_{3/2}$\cite{Johnson2008} using counter-propagating $780_0$ and 480 nm light. With $\sigma_+$ polarization for both beams we couple to  the Rydberg state $\ket{r}=\ket{nd_{5/2},m_j=5/2}$ which is selected with a $B_z=0.37~\rm  mT$ bias field. The other qubit ground state is $\ket{1}\equiv\ket{5s_{1/2},f=1,m_f=0}$. 
Coupling between  $\ket{\bar 1}, \ket{\bar r}$ is performed with $780_1$ and 480 nm light where $780_0$ and $780_1$ have the same propagation vector and polarization but a frequency difference of 6.8 GHz corresponding to the $^{87}$Rb $f=1\leftrightarrow f=2$ clock frequency.  In the experiments reported below we used Rydberg levels $97d_{5/2}$ and $111d_{5/2}$. 
In both cases strong blockade was observed in individual ensembles with 
no evidence for  double excitation of the logical $|\bar 1\rangle$ state\cite{Ebert2014}.  While we do not observe double excitation of 
$\ket{\bar 1}$, experiments with two ensembles do show  evidence for double excitation of the Rydberg state $\ket{\bar r}$, which plays a role in limiting the fidelity   with which we can prepare the $|\bar1\rangle$ state.

\begin{figure}[!t]
  \includegraphics[width=0.9\columnwidth]{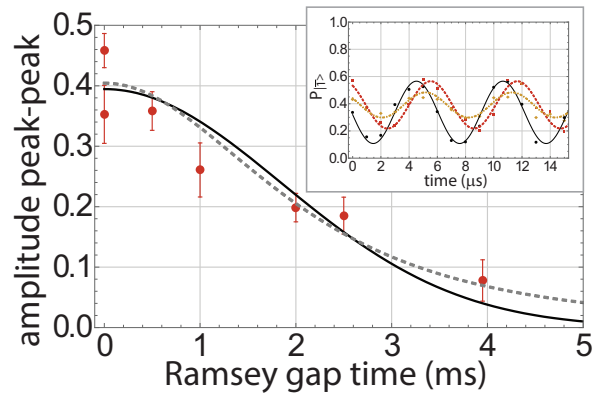}
\vspace{-.3cm}
  \caption{(color online) 
    Ramsey interference measurement of qubit coherence for $\bar N = 7.6$. The peak-peak amplitude of the oscillation
as a function of the gap time gives $T_2=2.6(3) ~\rm ms$. The circles are data points with $\pm \sigma$ error bars and the dashed and solid lines are fits to the functions $v_{\rm a}(t), v_{\rm b}(t) $ defined in the text. The gap time is the time $t$ between the $R_1(\pi)$ pulses in Eq. (\ref{eq.Ramsey}). All data have been corrected for  $\sim 1.5 \%$ probability per atom of the blow away giving an unwanted transition from $\ket{0}\rightarrow \ket{1}$.  The inset shows the Ramsey oscillations for
gap times of 0 (solid line), 0.5 ms (dashed line), and 2.5 ms (dashed-dotted line). 
  }
  \label{fig_WStateProc}
\end{figure}

We proceed to demonstrate the coherence of the ensemble states of Eq. (\ref{eqn_qubitBasis}) using Ramsey interferometry. The amplitude of the Ramsey signal is used to quantify the presence of $N$ atom entanglement in the ensemble, as has  been observed in other recent experiments\cite{Haas2014,Zeiher2015}. Details of the analysis showing that $82\pm6\%$ of the atoms  participate in the entangled $|W\rangle$ state are presented in the supplemental material\cite{Ebert2015SM}.  We load $3<\bar{N}<10$ atoms into one of the optical traps. The number of atoms loaded for each measurement follows a Poisson distribution with mean $\bar N$.  Each measurement starts with optical pumping into $\ket{\bar 0}$ followed by the pulse sequence 
\begin{equation}
 \ket{\psi}= R_1(\pi)R_0(\pi/2) R_1(\pi)G(t)R_1(\pi)R_0(\pi/2)\ket{\bar 0}.  
\label{eq.Ramsey}
\end{equation}
Here $R_0(\theta)$ is a pulse of area $\theta$ between states $\ket{\bar 0}, \ket{\bar r}$ and 
$R_1(\theta)$ is a pulse of area $\theta$ between states $\ket{\bar 1}, \ket{\bar r}$. The first $R_0(\pi/2)$ pulse creates an equal superposition 
$\frac{\ket{\bar 0}+\ket{\bar r}}{\sqrt2}$. This is then mapped to $\frac{\ket{\bar 0}+\ket{\bar 1}}{\sqrt2}$ with a $R_1(\pi) $ pulse, we wait a gap time $t$ described by an operator $G(t)$, map $\ket{\bar 1}\rightarrow \ket{\bar r}$ with a $R_1(\pi)$ pulse,   and then perform another $\pi/2$ pulse between $\ket{\bar 0}, \ket{\bar r}$. Finally, any population left in $\ket{\bar r}$ is mapped back to $\ket{\bar 1}$ with another $R_1(\pi)$ pulse. Atoms in state $\ket{0}$ are then pushed out of the trap using unbalanced radiation pressure from  a beam resonant with $\ket{5s_{1/2},f=2}\rightarrow \ket{5p_{3/2}, f=3}$ while the dipole trap light is chopped on and off. For the push out step a bias field is applied along $x$ the narrow axis of the dipole traps, and the circularly polarized push out beam propagates 
along $x$.   This is followed by a measurement of the number of atoms remaining in the dipole trap.    The resulting data are shown in Fig. \ref{fig_WStateProc}. The amplitude of the Ramsey interference at short gap times is limited by the $|W\rangle$ state preparation fidelity of about 50\% for the 
atom number used in the figure. 
 The fidelities of the $R_0(\pi)$ and $R_1(\pi)$ pulses used to prepare $|W\rangle$ are estimated to  each be  at least 90\% on the basis of previous experiments\cite{Ebert2014} and the strong inter-ensemble blockade effect we report below.  We attribute the limited $|W\rangle$ state preparation fidelity to Rydberg dephasing, as will be discussed in the following.  Periodic fluorescence measurements of the mean atom number (described in the supplemental material to \cite{Ebert2014}) bound  drifts to $6.7< \bar{N}< 9$, during the 12 hour measurement of this data set.

The principal sources of decoherence in this experiment are expected to be magnetic noise, motional dephasing, and 
atomic collisions\cite{Saffman2005a}. For small atom numbers and low collision rates we fit the Ramsey signal to the expression\cite{Kuhr2005}
$v_{\rm b}(t,T_{2})=v_0/[1+(e^{2/3}-1)(\frac{t}{T_{2}})^{2}]^{3/2}$ and in the collision dominated regime we use a Gaussian form $v_{\rm a}(t)=v_0 e^{-(t/T_2)^2}$ where $v_0$ is the amplitude at $t=0$. Both functional forms give the same $T_2$ time within our experimental error bars of   $T_2 = 2.6 \pm 0.3~\rm  ms$. 
The $\pi$ pulse times were $0.24~\mu\rm s$ for $\ket{\bar 0}\rightarrow \ket{\bar r}$, $0.06~\mu\rm s$ for the gap between pulses, and  $0.68~\mu\rm s$ for $\ket{\bar r}\rightarrow \ket{\bar 1}$ giving a  coherence to $R(\pi)$ gate time ratio of approximately $2600$.

\begin{figure}[!t]
  \includegraphics[width=0.9\linewidth]{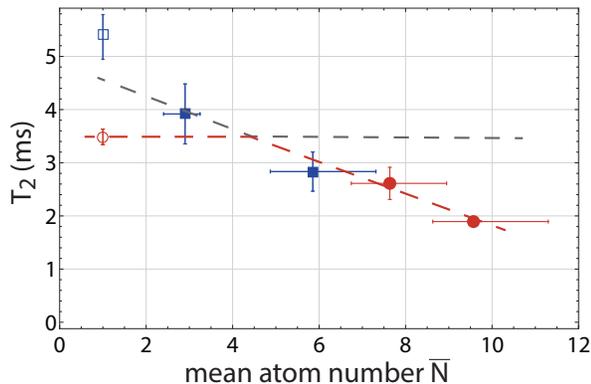}
\vspace{-.3cm}
  \caption{(color online)
   Dependence of ensemble coherence time on $\bar N$ for $\ket{W}$ states (red circles) and product states (blue squares). 
    The horizontal error bars represent the bounds for atom number measurements interleaved between Ramsey measurements. The open symbols are for preselected $N=1$ states. The dashed lines are  a guide to the eye.
}
  \label{fig_WStateT2}
\end{figure}

To further clarify the sensitivity to collisional dephasing Fig.  \ref{fig_WStateT2} shows the measured $T_2$ for different $\bar N$, including the case of $N=1$ Fock states which are selected using an additional fluorescence measurement before  the Ramsey sequence\cite{Ebert2014}. We see that $T_2\sim 1/{\bar N}$, in contrast to the $1/N^2$ scaling observed for GHZ states\cite{Monz2011}. 
The observed $1/\bar N$ scaling for $|W\rangle$ states is expected for decoherence dominated by collisions since the collision rate per atom is proportional to $\bar N$. 
For comparison, the $T_2$ time was also measured for product states $\ket{\psi}\sim (\ket{0}-i\ket{1})^{\otimes N}$. These states were prepared using a two-frequency Raman laser coupling $\ket{0}$ and $\ket{1}$ via the $5p_{3/2}$ level\cite{Yavuz2006} as shown in Fig. \ref{fig_maindia}. 
Comparison of the $\ket{\bar 1}$ ($\ket{W}$ state) and product state coherence data suggests that for $N \gtrsim 5$ the coherence time is limited by collisions. For $\bar N<5$ as well as for the $N=1$ Fock state data the product states show longer coherence time. The coherence of the $\ket{W}$ states is measured by comparison with a phase reference defined by the beatnote of the $780_0$ and $780_1$ Rydberg lasers which have a measured 
beatnote linewidth of 100 Hz FWHM. 
 This linewidth is consistent with the observed shorter coherence time of the $\ket{W}$ states compared to the product states which are referenced to the Raman laser beatnote which is in turn locked to a stable 6.8 GHz microwave oscillator. 
We anticipate that compensated optical traps and dynamical decoupling methods together with an optical lattice to reduce collisional effects can be used to greatly extend these coherence times\cite{Dudin2013}.

\begin{figure}[!t]
  \includegraphics[width=0.9\linewidth]{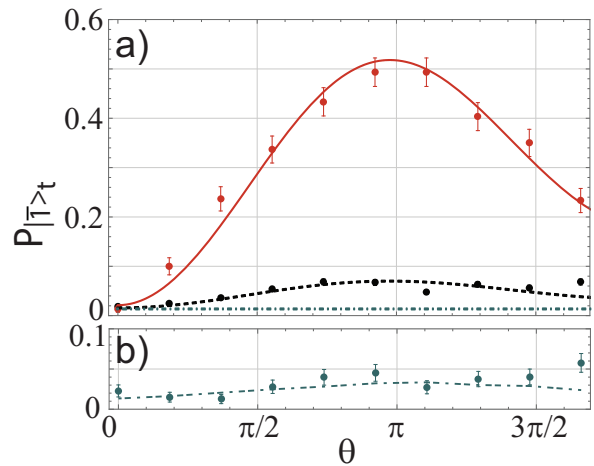}
\vspace{-.3cm}
  \caption{(color online)
Ensemble to ensemble blockade for $\bar N_{\rm c}= 9.9, \bar N_{\rm t}=6.2.$ 
    a) Probability of preparing $\ket{\bar 1}_{\rm t}$ without blockade (red circles, solid line) and with blockade (black circles, dashed line). 
The solid line is a fit to a decaying sinusoid function from \cite{Ebert2014}. 
    The dashed line is the same fit scaled by 11\%.
    b) Blockade data post selected on detection of $\ket{\bar 1}_{\rm c}$.  The dashed-dotted lines in both panels show the expected signal due to state leakage during blow-away in the control and target regions.
  }
  \label{fig_EEBTimeSeries}
\end{figure}

To demonstrate ensemble-ensemble blockade we load atoms into control (c) and target (t) dipole traps, optically pump into $\ket{\bar 0}_c\ket{\bar 0 }_t$ and apply one of two sequences. Preparation of a superposition of $\ket{\bar 0}$ and $\ket{\bar 1}$ in the target qubit is effected by the sequence $U_a\ket{\bar 0}_{\rm c}\ket{\bar 0}_{\rm t}=R_{1,\rm t}(\pi)R_{0,\rm t}(\theta)\ket{\bar 0}_{\rm c}\ket{\bar 0}_{\rm t}$. This should ideally leave the qubits in the joint state $\ket{\bar 0}_{\rm c}\left[\cos(\theta/2)\ket{\bar 0}_{\rm t}-\sin(\theta/2)\ket{\bar 1}_{\rm t}\right]$ with the probability of preparing $\ket{\bar 1}_t$ proportional to $\sin^2(\theta/2),$ as is shown in Fig. \ref{fig_EEBTimeSeries}a). We see the expected time dependence with a peak probability of $P_{\ket{\bar 1},\rm t}\sim 0.52$, consistent with our earlier study of Fock state preparation\cite{Ebert2014}. 

Rydberg blockade between two ensembles is observed with the sequence $U_b\ket{\bar 0}_{\rm c}\ket{\bar 0}_{\rm t}=R_{1,c}(\pi)R_{1,t}(\pi)R_{0,t}(\theta)R_{0,c}(\pi)\ket{\bar 0}_{\rm c}\ket{\bar 0}_{\rm t}$. Here we have used state $\ket{\bar 0}$ of the control ensemble to block the target transfer with the final $R_{1,c}(\pi)$ pulse 
ideally leaving the qubits in the joint state $\ket{\bar 1}_{\rm c}\ket{\bar 0}_{\rm t}$. 
The data in Fig. \ref{fig_EEBTimeSeries}a) show a ratio of $P_{\ket{\bar 1},t}(U_b)/P_{\ket{\bar 1},t}(U_a)=0.11(1)$, i.e. a blockade fidelity of 0.89.  
This implies that  the success probability of the transition $R_{0,\rm c}(\pi) \ket{\bar 0}_{\rm c}\rightarrow \ket{\bar r}_c$ is bounded below by  the  $\ket{\bar 1}_{\rm t}$ population  ratio for the two sequences. We infer that at least one atom is excited to the Rydberg  state $\ket{r}_{\rm c}$ with probability  $\ge 0.89(1)$.

As a further check on the inter-site blockade fidelity, events where the control site ends in state $\ket{\bar 1}_{\rm c}$ after sequence $U_b$ are post selected. 
The observed post-selected target population is shown in Figure \ref{fig_EEBTimeSeries}b), along with the expected blow-away leakage rate of the control and target sites which is measured to be $0.2\%/\rm atom$. 
From the data it can be seen that the post-selected results are consistent with perfect inter-site blockade. 

\begin{figure}[!t]
  \includegraphics[width=0.99\columnwidth]{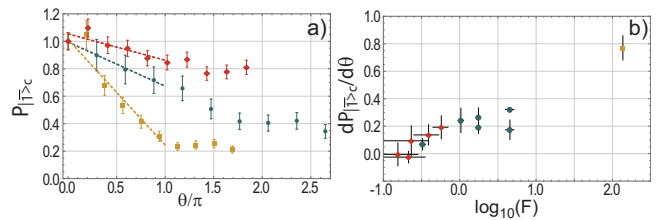}
\vspace{-.7cm}
  \caption{(color online) Probability of preparing state $\ket{\bar 1}_{\rm c}$ as a function of the target ensemble pulse area $\theta$. 
    a) Probability for several parameter sets: 
$(111d_{5/2}$, $R=8.3$ and $8.7~\mu\rm m)$ (red diamonds),
 $(97d_{5/2}$, $R=8.3$ and $8.7~\mu\rm m)$ (green circles),
$(97d_{5/2}$, $R=17~\mu\rm m)$ (yellow squares). The data has been normalized to 1 at $\theta=0$ for clarity, with typical success probability  40-60\%.
    b) Comparison of the slope of the data in panel (a) with the scaling  parameter $F$ from Eq. (\ref{eqn_testparam}).
   The color markers are the same as in panel a). 
  }
  \label{fig_CTR}
\end{figure}

The observed high blockade fidelity exceeds that originally achieved in experiments with single atom qubits\cite{Gaetan2009,Urban2009}, and is certainly sufficient to create entanglement between ensemble qubits. What has so far limited a demonstration of deterministic entanglement is the relatively low probability of up to 62\% \cite{Ebert2014} with which the ensemble state $\ket{\bar 1}$ can be prepared. In order to gain insight into what is limiting the state preparation fidelity we looked for signatures of Rydberg-Rydberg interactions concurrently with strong blockade. Ideally the probability of preparing $\ket{\bar 1}_{\rm c}$ with sequence $U_b$, should be independent of the pulse area $\theta$ applied to the target ensemble.  However a clear dependence on $\theta$ can be seen in Fig.  \ref{fig_CTR}a).
We believe this effect is due to long range interactions, where the amplitude for Rydberg atom excitation in the target site is sufficiently blockaded to prevent it from making the transfer to $\ket{\bar 1}_{\rm t}$ with any significant probability, yet the target ensemble Rydberg excitation still interacts with the control ensemble strongly enough to disrupt the control ensemble state transfer. A similar situation of partial blockade together with decoherence of multi-atom ground-Rydberg Rabi oscillations was reported earlier in \cite{Johnson2008}.

A two-atom Rydberg interaction  effect should scale with the Rydberg double excitation probability,  i.e. $P_2  \propto \Omega_{\bar{N}}^2/{\sf B}^2$, where $\sf B$ is the ensemble mean blockade shift\cite{Walker2008}.
To check this, we extract the slopes from linear fits to the $P_{\ket{\bar 1}_{\rm c}}(\theta)$ data for small $\theta$ and compare to the 
 scaling  parameter 
\begin{equation}
  \label{eqn_testparam}
  F = \Omega_{\bar{N}_{\rm t}}^2 \left[\frac{(n/n_0)^{12}}{(R/R_0)^6}\right]^{-2} \propto P_{\rm double}.
\end{equation}
Here $n$ is the Rydberg principal quantum number and $R$ is the site - site separation. 
The larger $F$ is for a given set of parameters, the stronger the  Rydberg-Rydberg interaction, and thus the larger the slope of  
$d P_{\ket{\bar 1}_{\rm c}}(\theta)/d\theta.$
Indeed, this is the behavior we observe, as shown in Fig. \ref{fig_CTR}b), for a range of $\bar{N}$,  $R$, and $n$.

This interaction effect hints at the possible mechanism responsible for the observed reduction in the probability $P_{\ket{\bar 1}}$ of preparing the collective qubit state in a single ensemble. The spatial extent of one ensemble is $\sim 2\sigma_z=14~\mu\rm m$ giving a length scale in between the lower two data sets in Fig. \ref{fig_CTR}a). The intra-ensemble Rydberg interactions are significantly stronger than between atoms located in different ensembles at the same separation because the dipole-dipole interaction angular factors favor atom pairs separated along $z$\cite{Walker2008}. These considerations imply that  lack of perfect blockade  leading to long range Rydberg-Rydberg interactions in a single ensemble only partially explains 
the observed maximum of $P_{\ket{\bar 1}}=0.62$ \cite{Ebert2014}. Another candidate explanation is very strong interactions at short range in a 
single ensemble  which mix  levels together and open anti-blockade resonance channels\cite{Amthor2010}. The doubly excited  molecular energy structure becomes difficult to calculate with confidence at short range, with many molecular potentials near resonant\cite{Schwettmann2006,*Keating2013}.
For our typical Rydberg state $97d_{5/2}$ this characteristic separation is $\sim 5$ $\mu$m, and for a 6 atom sample with our ensemble spatial distributions an average of 7 atom pairs out of 15 have $R< 5~\mu\rm m$. We conjecture that the  strong, short range interactions 
give an  amplitude for double excitation, resulting in Rydberg-Rydberg interactions which dephase the ground-Rydberg rotations needed for state preparation, thereby   limiting the probability of preparing the ensemble $\ket{\bar 1}$ state. A related reduction of the fidelity of Rydberg mediated  atom-photon coupling  in dense ensembles due to Rydberg-ground state interactions has also been observed\cite{Baur2014}. 

In conclusion, we have demonstrated the coherence of  ensemble qubit basis states. The coherence time scales approximately inversely with the number of atoms, but is still several ms and $2600$ times  longer than our characteristic gate time for $N\sim 10$. 
Additionally we have demonstrated inter-ensemble blockade with a fidelity of 0.89 and $\sim 1.0$ when post-selecting on control ensemble excitation. We identified Rydberg-Rydberg interactions from weak double excitations, either at long or short range,  as a possible mechanism limiting the fidelity of ensemble state preparation. Future work towards ensemble entanglement and quantum computation will explore the use of a background optical  lattice to better localize the ensembles while limiting  uncontrolled short range interactions.

 This work was funded by NSF grant PHY-1104531 and the AFOSR Quantum Memories MURI.

\bibliographystyle{apsrev4-1}

%\bibliography{d:/users/mark/pubs/biblio/saffman-refs,d:/users/mark/pubs/biblio/rydberg_bib_v15,d:/users/mark/pubs/biblio/qc_refs%,d:/users/mark/pubs/biblio/atomic}

\begin{thebibliography}{44}
\bibitem [{201()}]{2015_ensemble_note1}%
  \BibitemOpen
  \href@noop {} {}\bibinfo {note} {Equation (\ref{eqn_qubitBasis}) is an
  approximation to $\ket{\bar 1} = \sum_{j=1}^N\frac{\Omega_{j}}{\Omega_N}
  |0_{1}0_{2}... 1_j ... 0_{N}\rangle$ with $\Omega_j$ the Rabi frequency of
  the $j^{\rm th}$ atom. For our experiments the Rabi frequency is
  approximately constant across the ensemble so $\Omega_N=N^{1/2} \Omega$ and
  Eq. (\ref{eqn_qubitBasis}) can be written as $\ket{\bar 1} =
  N^{-1/2}\sum_{j=1}^{N}|0_{1}0_{2}... 1_{j} ... 0_{N}\rangle$.}\BibitemShut
  {Stop}%
\end{thebibliography}

%merlin.mbs apsrev4-1.bst 2010-07-25 4.21a (PWD, AO, DPC) hacked
%Control: key (0)
%Control: author (72) initials jnrlst
%Control: editor formatted (1) identically to author
%Control: production of article title (-1) disabled
%Control: page (0) single
%Control: year (1) truncated
%Control: production of eprint (0) enabled
%

%%%%%%%%%%%
% Supplemental material
%%%%%%%%%%%%

\newpage

{\bf Coherence and Rydberg blockade of atomic ensemble qubits\\Supplementary Material}

\section{Multipartite W-State Entanglement Verification}

In order to demonstrate multipartite entanglement it is necessary to show that the results obtained in a measurement cannot be reproduced with a separable state.
Thus we require that  the $N$-particle state in question $|\psi^{N}\rangle$ satisfies 
\begin{equation}
  |\psi^{N}\rangle \neq |\psi_A^{K}\rangle \otimes |\psi_B^{N-K}\rangle,
\end{equation}
for any  K in the range $N/2\leq K<N$.
In this supplemental material we evaluate the observed signatures of $W$-state entanglement.
These signatures include the $\sqrt{\bar{N}}$-enhancement of the Rabi frequency between $|\bar{0}\rangle$ and $|\bar r\rangle$, and the amplitude of the Ramsey oscillations.

\subsection{Collective Rabi Frequency Enhancement}

The interaction of an ensemble with a light field can be written in the basis of individual atom excitations $ |\{0,1\}^{(1)}\rangle \otimes |\{0,1\}^{(2)}\rangle \otimes ... |\{0,1\}^{(N)}\rangle$.
The Hamiltonian $\mathbf{H}_{int}$ describing the evolution of the system is a block tridiagonal matrix. The basis states 
 are denoted as $|n_k\rangle$, where $0\le n\le N$ is the eigenvalue of the excitation number operator $\hat{\mathcal N}= \sum_{k=1}^N \hat{S}_z^{(k)} + N/2,$ and the index $k$ labels the  degenerate eigenstates, e.g. $|1_1\rangle=|10\cdots 0\rangle,$ $|1_2\rangle=|01\cdots 0\rangle$, etc. .
Here $\hat{S}_{z}^{(k)}=\frac{1}{2}\hat\sigma_z^{(k)}$ is the effective spin operator for atom $k$ along $z$.
In this basis $\mathbf{H}_{int}$ is given by:
\begin{equation}
  \label{eqn_blocktridiagonal}
  \begin{split}
  &\mathbf{H}_{int} = \mathbf{A} + \mathbf{\Delta} =\\
  &\begin{bmatrix}
    \mathbf{\Delta}_{0}  & \mathbf{A}_{(0,1)}  & 0 & \cdots & 0 \\
    \mathbf{A}^{T}_{(0,1)}  & \mathbf{\Delta}_{1} & \mathbf{A}_{(1,2)} & & \\
    \vdots & \ddots & \ddots  & \ddots  & \vdots \\
           &        & \mathbf{A}^{T}_{(N-2,N-1)} & \mathbf{\Delta}_{N-1} & \mathbf{A}_{(N-1,N)}   \\
    0      & \cdots & 0 & \mathbf{A}^{T}_{(N-1,N)}   & \mathbf{\Delta}_{N}
  \end{bmatrix}
  \end{split}
\end{equation}
The matrix $\mathbf{H}_{int}$ has dimensions $2^N\times 2^N$, the dimension of $N$ 2-level systems.
The dimension of the block diagonal sub-matrices is given by the binomial coefficient, $\textrm{dim}\left( \mathbf{\Delta}_n \right) = \tbinom N{n} \equiv N_{n}$.
The sub-matrices, $\mathbf{\Delta}_{n}$, contain information concerning the sub-systems specific energy levels
\begin{equation}
  \label{eqn_bdsubmatricieselem}
\mathbf{\Delta}_{n} = \sum_{k=1}^{N_{n}}\delta^{(n)}_k |n_k\rangle\langle n_k|
\end{equation}
where $\delta^{(n)}_k$ refers to the energy of the $k^{\mathrm{th}}$ basis state in the subspace with eigenvalue $n$.

The matrices on the upper and lower diagonals  couple states with excitation numbers differing by $\pm1$, 
 $|n_k\rangle \stackrel{\alpha}{\leftrightarrow} |n\pm 1_j\rangle$ with coupling strength $\alpha$ defined by
\begin{eqnarray}
  \label{eqn_couplingsubmatricies}
\left[\mathbf{A}_{(n\pm 1,n)}\right]_{j,k}&=& \alpha_{jk}|(n\pm1)_j\rangle\langle n_k| \nonumber\\
& =& |(n\pm1)_j\rangle\langle (n\pm1)_j|\hat{A}|n_k\rangle\langle n_k|,
\end{eqnarray}
where
\begin{equation}
  \label{eqn_aop}
  \hat{A} = \sum_{m=1}^{N} \alpha_m  \hat{S}_x^{(m)}
\end{equation}
and $\alpha_m$ is the strength of the light-atom coupling at atom $m$. 
In an ideal Rydberg blockaded ensemble states with $n>1$ are not excited and all $\alpha_m$ are equal. Departures from the ideal case are accounted for by allowing for atom specific $\alpha_m$ and double excitations are included by truncating the basis at $n=2$ and adding the doubly excited interaction energies to $\mathbf\Delta_2$.  

A strong blockade shift, $\delta_m^{(n=2)} = \delta_{dd} \gg \alpha_m$ reduces the available Hilbert space for the problem to $n=\{ 0,1 \}$, and $\mathbf{H}_{int}$ becomes:
\begin{equation}
  \label{eqn_hintsimp}
  \mathbf{H}_{int} = \begin{bmatrix}
    0 & \alpha_1 & \alpha_2 & \cdots & \alpha_N \\
    \alpha_1 & \delta_1^{(1)} & 0 & \cdots & 0 \\
    \alpha_2 & 0 & \delta_2^{(1)} &  & 0 \\
    \vdots & \vdots & & \ddots  & \vdots \\
    \alpha_N & 0 & 0 & \cdots & \delta_N^{(1)} 
  \end{bmatrix}
\end{equation}
The detunings $\delta_m^{(1)}$ are nominally 0, so it makes sense to treat the $\delta^{(1)}$ entries as a perturbation.
Under the condition of perfect blockade and no detuning, the energy eigenstates of $ \mathbf{H}_{int}=\mathbf{A}$ are the dressed states 
$\frac{1}{\sqrt{2}}\left(|\bar{0}\rangle \pm |\bar{1}\rangle\right)$ with total angular momentum $J=N/2$ and $N-1$ orthogonal states with total angular momentum $J = (N/2 -1)$: $\{\frac{1}{\sqrt{2}}\left(|\bar{0}\rangle \pm |\bar{1}\rangle\right), |(\bar{1})_{\perp}\rangle\}$, where $|\bar{1}\rangle \equiv \sum_{k=1}^N \frac{\alpha_k}{\bar \alpha_N}|1_k\rangle$ with $\bar\alpha_N^2 \equiv \sum_{k=1}^N \alpha_k^2$.
The eigenvalues determine the speed at which the system evolves, for $\frac{1}{\sqrt{2}}\left(|\bar{0}\rangle \pm |\bar{1}\rangle\right)$ this speed is $\pm \alpha_N$ implying a collective enhancement of $\sqrt{N}\alpha$ when the coupling strengths are homogeneous.

\begin{figure}[!t]
\includegraphics[width=0.45\textwidth]{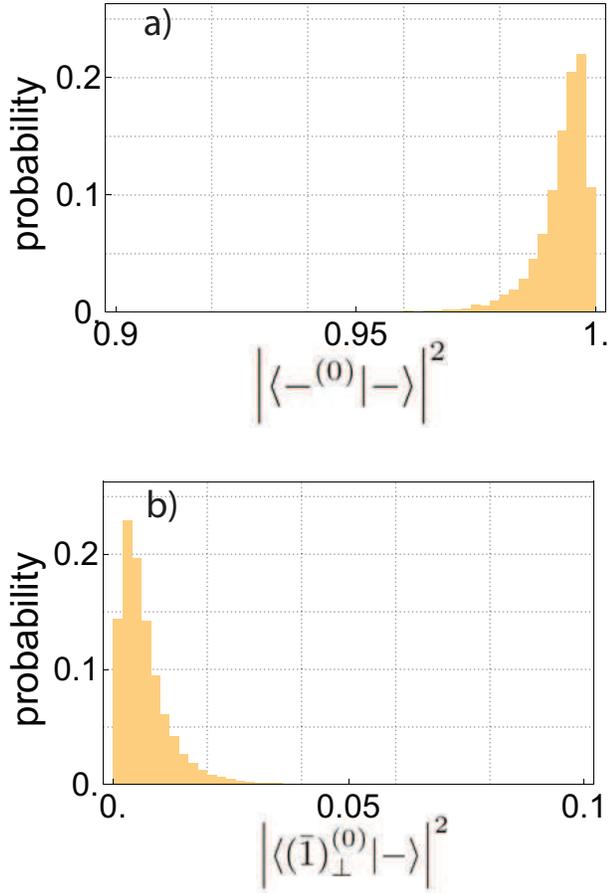}
  \caption{
  \label{fig_eignstateproj}
    Monte Carlo calculations for $N=5$ atoms for 10,000 randomized instances of atom positions and velocities consistent with our experimental parameters  of (a) Projection of a symmetric eigenstate without inhomogeneous broadening,
 $|-^{(0)}\rangle = \frac{1}{\sqrt{2}}\left(|\bar{0}\rangle - |\bar{1}\rangle\right)$, onto the energy eigenstate of the full inhomogeneous Hamiltonian $\mathbf{H}_{int}|-\rangle = E_-|-\rangle$ and  
    (b) projection of the symmetric eigenstate onto the orthogonal subspace  $\{|(\bar{1})_{\perp}^{(0)}\rangle\}$.
   }
\end{figure}

Our system has low inhomogeneous coupling contributions as evidenced by the $\alpha_N = 0.96\sqrt{\bar{N}}\alpha$ scaling observed in our previous work \cite{Ebert2014}, for reference an average scaling of $0.972$ is predicted from experimental parameters.
The observation of $\sqrt{\bar N}$ scaling of the coupling strength is a classic signature of Rydberg blockade and $N$ participating wavefunctions, as the $|\bar{1}\rangle$ state is the only state that can evolve with that coupling strength.
A state with $k$-partite entanglement consistent with the observed perfect blockade given by, $|\psi^{N}\rangle = |\bar{1}^{k}\rangle \otimes |\bar{0}^{N-k}\rangle$, will still oscillate at the same $\sqrt{\bar{N}}$ frequency, but the amplitude will be reduced to the overlap with $|\bar{1}^{N}\rangle$, $|\langle \bar{1}^N |\psi^{N}\rangle |^2 = k/N$, this is discussed further in the next section.

The orthogonal singly-excited states $|(\bar{1})_{\perp}\rangle$ do not couple to the symmetric states $\{|\bar{0}\rangle,|\bar{1}\rangle\}$ under ideal conditions ($\delta_k^{(1)}=0$).
This becomes clear when the Bloch picture is invoked, since the symmetric states have total angular momentum $J=N/2$ while the $|(\bar{1})_{\perp}\rangle$ states have $J=N/2-1$ and a rotation on the Bloch sphere conserves angular momentum.
Inhomogeneous broadening, including differential AC Stark shifts, Doppler shifts, and finite intermediate state lifetimes, are added perturbatively with $\mathbf{\Delta}$ and provide a mechanism for coupling into the $|(\bar{1})_{\perp}\rangle$ space.
This coupling should be negligible and reduce with increasing $N$ and additionally will not display the characteristic $\sqrt{N}$ enhancement.
Figure \ref{fig_eignstateproj} shows simulated projections of $\frac{1}{\sqrt{2}}\left(|\bar{0}\rangle - |\bar{1}\rangle\right)$ along the energy eigenstates of $\mathbf{H}_{int}$ for $N=5$ atoms with our experimental parameters.

\subsection{Coherence Amplitude}

Since the coupling to the orthogonal subspace is negligible for our experimental parameters, the amplitude of the Ramsey fringe oscillations provide a threshold for entanglement.
A thermal sample of singly excited states $|1_{\rm th}\rangle=\sum_{k=1}^{N} e^{\imath\phi_k}|1_k\rangle$, where $\phi_k$ is a random phase factor for the $k^{\mathrm{th}}$ atom, will only couple back to $|\bar{0}\rangle$ by the amount of overlap with the $|\bar{1}\rangle$ state.
The projection $|\langle \bar{1}|1_{\rm th}\rangle|^2$ will average to $1/N$, therefore an oscillation with contrast above $1/N$ cannot be a thermal sample.

\begin{figure}[!t]
 \includegraphics[width=0.45\textwidth]{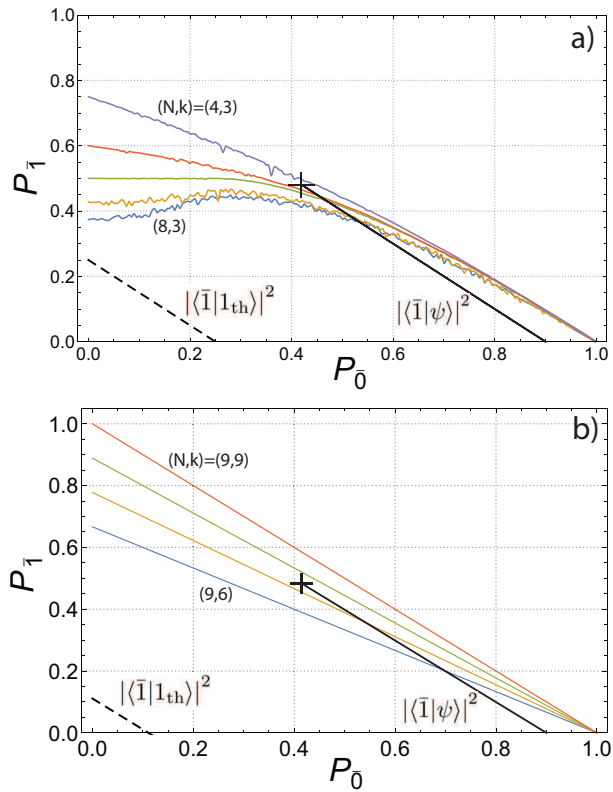}
  \caption{(color online)
    a) Numerically determined bounds for $(P_{\bar 0}, P_{\bar{1}})$ using  Eq. (\ref{eq.psient}). Rydberg blockade is not assumed in the calculation so multiple excitations are allowed. 
    States above the $(N,k)$ line imply there are at least $k$ entangled atoms in the $N$ atom  ensemble. 
    Calculated bounds for $N=4-8$ are shown, top to bottom. The dashed black line shows the amplitudes  for the thermal singly-excited state $|1_{\rm th}\rangle$ with $N=4$.
    The solid black line represents the range of experimental Ramsey oscillation data with the cross showing the value at  $t_{\rm gap} =0$ ms from Fig. 2 in the main text. b)  Analytical bounds assuming perfect blockade using Eqs. (\ref{eq.psientB},\ref{eqn_threshold}). The entanglement thresholds are the straight lines shown for  $N=9$ and $k=9-6,$ top to bottom. The data shown by the black line and cross exceeds the $k=7$ threshold.
   \label{fig_ReichelKpartiteNum} }
\end{figure}

To generate a threshold for $k$-partite entanglement we perform a numerical simulation along the lines of the analysis   in   \cite{Haas2014}.
Briefly, the goal is to generate an upper bound on a measurement of $P_{\bar{1}}=|\langle \bar{1}|\psi\rangle|^2$ as a function of $P_{\bar 0}=|\langle\bar{0}|\psi\rangle|^2$ for states $|\psi\rangle$ with a maximum of $k$ entangled particles. We establish bounds in two ways. First, we do not assume Rydberg blocakde so multiple excitations are possible. 
This is done by creating a random $k$-partite entangled wavefunction 
\begin{equation}
|\psi\rangle = |\psi^{(k)}_1\rangle \otimes  ... |\psi^{(k)}_{m-1}\rangle \otimes |\psi^{(k_m)}_m\rangle,
\label{eq.psient}
\end{equation}
 where $|\psi^{(k)}_i\rangle = \sin \left( \theta_i/2\right) |\bar{0}^{(k)}\rangle+\cos \left(\theta_i/2\right)e^{i\phi_i} |\bar{1}^{(k)}\rangle$, $\theta_i$ and $\phi_i$ are randomly generated, and $k_m = N-(m-1)k$.
We extract the maximum $P_{\bar{1}}$ for a given $P_{\bar 0}$ bin obtained numerically to arrive at the thresholds shown in Fig. \ref{fig_ReichelKpartiteNum}a) for $k=3$ particle entanglement with ensemble atom numbers $N=4-8$.
Any state above the threshold must have at least $k$-partite entanglement. The black cross is an experimental data point recorded for a sample with $\bar N=8.8$ atoms, verifying the presence of entanglement.

Rydberg blockade limits the Hilbert space to $n\leq 1$ excitations, which simplifies the calculation and enables an analytical bound  for the $k$-partite entanglement threshold.
The state in Eq. (\ref{eq.psient})  includes kets with multiple excitations. To remove these  we impose the blockade condition $P_{(n>1)}=0$ and write the state as 
\begin{equation}
|\psi\rangle=\left(a_1|\bar{0}^{(k)}\rangle+b_1 |\bar{1}^{(k)}\rangle\right) \otimes |\bar{0}^{(N-k)}\rangle.
\label{eq.psientB}
\end{equation}
Maximization of $P_{\bar{1}}$ for a given $P_{\bar 0}$ can be readily accomplished analytically to give
\begin{equation}
  \label{eqn_threshold}
  P_{\bar{1}} = \frac{k}{N}(1-P_{\bar 0}).
\end{equation}
Note that this agrees with the limiting case of $P_{\bar 0}=0$ from \cite{Haas2014}.
Rearranging (\ref{eqn_threshold}) to give $\frac{k}{N} \leq \frac{P_{\bar{1}}^{\mathrm{max}}}{1-P_{\bar 0}^{\mathrm{max}}}$, and given our  extreme value $(P_{\bar0},P_{\bar{1}})=(0.44\pm0.02,0.46\pm0.03)$ we can show that we meet the threshold for creation of the $W$-state with $\frac{k}{N}=82\pm 6\%$. Similar arguments for the presence of entanglement based on the amplitude of Ramsey oscillations have been used in \cite{Zeiher2015}.\\

\section{Summary}

In summary we have shown evidence for $N$ particle W-state entanglement on the basis of the following three arguments.
First, the excellent agreement of the observed collectively enhanced Rabi frequency with 
theory reported in our previous work\cite{Ebert2014} using the same experimental setup and procedures as are used here implies an $N$ component wavefunction.
Second, the amplitude of the Ramsey-style oscillations for $\bar N=8.8$ is four times larger than the $1/\bar N$ limit expected from a thermal sample of singly excited states. Our data shows entanglement without making the assumption of perfect blockade.  
Third, with the assumptions of perfect blockade, entanglement percentage independent of $N$, and negligible coupling to $\{|(\bar 1)_\perp\rangle \}$, which is justified by Fig. 1,  then  $\frac{k}{N}=82\pm 6\%$. In other words 
$82\pm 6\%$ of the atoms in the ensemble are participating in the $W$-state entanglement. This result is not changed in a statistically significant way when compared with  simulations based on experimental parameters that include imperfect blockade.

\end{document}